\documentclass[reprint,twocolumn,amsmath,amssymb,aps,pra,longbibliography,floatfix]{revtex4-1}
\usepackage{physics}
\usepackage{amsmath}
\usepackage{cases}
\usepackage{url}
\usepackage{natbib}
\usepackage{textcase}
\usepackage{amssymb}
\usepackage{graphicx}
\usepackage{bm} 
\usepackage{times}
\usepackage{float}
\usepackage{multirow,microtype,color,relsize}
\usepackage{subfigure}
\usepackage[breaklinks=true]{hyperref}
\usepackage[utf8]{inputenc}
\usepackage[english]{babel}
\hypersetup{colorlinks=true,linkcolor=blue,citecolor=blue}
\hypersetup{linktocpage}
\usepackage{CJKutf8}
\usepackage{breakcites}
\usepackage{float}
\usepackage[dvipsnames]{xcolor}
\definecolor{mypine}{RGB}{1, 121, 111}

\begin{document}
\title{Bound state-continuum resonance transition in a shallow quantum well}
\author{Yi Huang}
\author{Sankar Das Sarma}
\affiliation{Condensed Matter Theory Center and Joint Quantum Institute, Department of Physics, University of Maryland, College Park, Maryland 20742, USA}

\begin{abstract}	
We show a transition from a bound state to a continuum resonance in a shallow quantum well (QW) by electrostatic gating to bend the conduction band edge.
This bound state-continuum resonance (BSCR) transition is particularly relevant in topological quantum computing platforms where shallow InAs QWs are used. 
We predict the observed capacitance jump and the parallel metal-insulator transition accompanied the BSCR transition.
An experimental puzzle in shallow InAs QWs is the mobility drop at an electron density smaller than expected for the bound-state second subband occupation.
We explain this puzzle as a result of intersubband scattering involving a level-broadened continuum resonance, mediated by screened Coulomb impurities.

\end{abstract}
\maketitle

\begin{figure}[t]
    \centering \includegraphics[width=\linewidth]{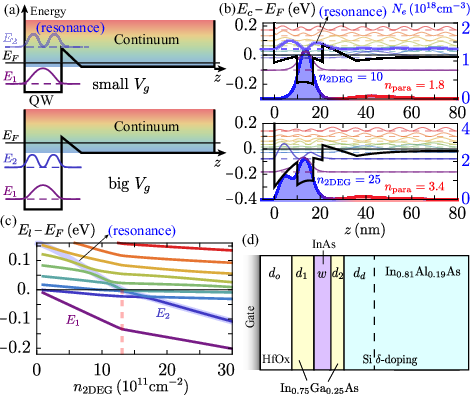}
    \caption{(a) Schematics of a bound state ($E_2$) hidden in the continuum above a shallow quantum well (QW) at small gate voltage $V_g$. (b) The conduction band edge $E_c$ (black) and the 3D electron density profile $N_e$ (blue and red) for a realistic shallow InAs QW (Sample B in Ref.~\cite{Manfra_100_InAs:2023}). The thin dashed (solid) colored lines represent different subband energies (wavefunctions).  (c) Subband energies $E_l$ versus $n_{\mathrm{2DEG}}$. The shaded blue line tracing the transition from a QW bound state to a continuum resonnance. (d) A schematic of the stack structure of a realistic modulation-doped InAs QW~\cite{Manfra_100_InAs:2023}.}
    \label{fig:quantum_well}
    \vspace{-0.2 in}
\end{figure}

\textit{Introduction}.---In quantum mechanics, a bound state is a spatially localized state due to potential confinement. Above the potential well, there is a continuum where plane-wave-like states extend the entire space, creating a continuous spectral range. Within the continuum, certain level-broadened energies, known as resonances, can develop scattering extrema, mimicking ``virtual bound states'' with finite lifetimes. 
One method to control bound states in low-dimensional electronic systems is through band-structure engineering in semiconductor quantum wells (QWs). 
This has been a major topic in semiconductor physics for the past half-century~\cite{Ando_review:1982}.
By engineering the potential depth of a QW, the energy levels of bound states and resonances can be shifted relative to the continuum bottom, allowing a bound state to transition to a resonance [cf. Fig.~\ref{fig:quantum_well} (a)]. 
This bound state-continuum resonance (BSCR) transition may manifest in transport experiments as a mobility drop when the Fermi level starts occupying the continuum resonance~\cite{Manfra_100_InAs:2023}. 
However, experimentally changing the potential depth continuously within the same device by altering material chemical composition is challenging. 
This difficulty can be overcome by using electrostatic gating, where the potential bending by gate electric field provides a continuous tunability.
The BSCR transition is easier to observe in a shallow QW, where only the lowest subband exists as a bound state while all excited subbands appear as resonances in the continuum.

A shallow InAs QW is an ideal material for observing the BSCR transition due to its small longitudinal effective electron mass $m=0.023 m_0$ (compared to $0.067 m_0$ in GaAs and $0.98 m_0$ in Si, where $m_0$ is the free electron mass), so that the large kinetic energy squeezes the excited subbands into the continuum above QW.
This makes shallow InAs QWs very different from the record-high-mobility GaAs~\cite{Pfeiffer:2021,Ahn_high_mobility_GaAs:2022,Huang:2022a} or Si~\cite{Esposti:2023,huang2023understanding} QWs, where 3-4 bound-state subbands are inside QWs.
Because of controlled proximity coupling enhanced by near-surface electrons, shallow InAs QWs hybrid with superconductors are considered prime candidates for fault-tolerant topological quantum computation~\cite{microsoft_InAs-Al_protocol:2023,microsoft_InAs-Al_interferometric:2024,Shabani:2021,Dartiailh:2021,Marcus:2023}. 
The key role of disorder in such InAs-based  topological quantum computing platforms has been extensively discussed in the literature~\cite{Ahn_disorder_Majorana:2021,Stanescu:2023,Haining_disorder_Majorana:2023,Das_Sarma_majorana:2023}. 
Understanding BSCR physics helps better comprehend fundamental physical properties such as mobility and capacitance that provides guidelines for enhancing the electrostatic control and sample quality of shallow InAs QWs and paves the way to topological quantum computation.

\begin{figure}[t]
    \centering  \includegraphics[width=\linewidth]{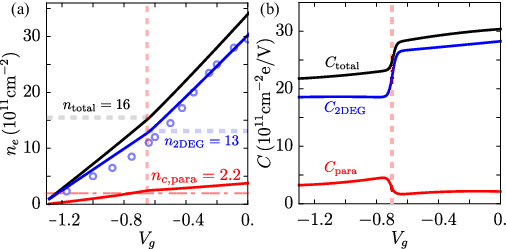}
    \caption{(a) Theoretically calculated electron densities $n_{\mathrm{total}}$ (black), $n_{\mathrm{2DEG}}$ (blue), and $n_{\mathrm{para}}$ (red) versus the gate voltage $V_g$. The blue circles represent the experimental 2DEG density $n_{\mathrm{2DEG}}$ measured in Ref.~\cite{Manfra_100_InAs:2023}. The dashed lines mark the density at which the second subband is occupied. The red dot-dashed line marks the parallel MIT critical density. (b) The calculated capacitance $C = d n_e/ d V_g$.}
   \label{fig:density_voltage}
    \vspace{-0.2 in}
\end{figure}

\begin{figure}[t]
    \centering
    \includegraphics[width = \linewidth]{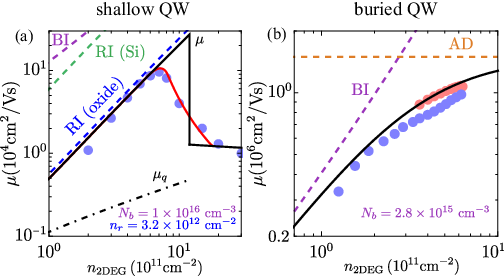}
    \caption{2DEG mobility $\mu$ vs density $n_{\mathrm{2DEG}}$ for (a) shallow InAs QWs in Ref.~\cite{Manfra_100_InAs:2023} and (b) buried QWs in Ref.~\cite{Hatke:2017}. Colored dots represent experimental data. The solid black (red) curve in (a) represents the theoretical total mobility without (with) level-broadening effect. Dashed curves represent the mobility individually contributed from different scattering mechanisms. The best-fit impurity densities are shown as insets.}
    \label{fig:mobility_density}
    \vspace{-0.2 in}
\end{figure}

In this letter, we demonstrate a BSCR transition with profound consequences for the quality of shallow two-dimensional (2D) InAs QWs.
This study is motivated by low-temperature transport measurements in state-of-the-art shallow InAs QWs reported in Ref.~\cite{Manfra_100_InAs:2023} [cf. Fig.~\ref{fig:quantum_well} (d)], which serve as the fundamental building blocks for topological quantum computing hardware~\cite{microsoft_InAs-Al_protocol:2023,microsoft_InAs-Al_interferometric:2024}.
The energy levels and the electron populations of the QW bound states and the continuum can be continuously tunned by applying a gate voltage.
At a low gate voltage, the potential well is shallow, with only the first subband occupied below the Fermi level ($E_1 < E_F$) as a bound state inside the QW, while the excited resonances reside in the continuum. 
Increasing the gate voltage deepens the potential well, populating more 2D electron gas (2DEG) inside the QW. 
Consequently, the resonance second subband is gradually pulled down from the continuum, eventually becomes a bound state [cf. Fig.\ref{fig:quantum_well} (a-c)].
If and when this second level is occupied, the mobility drops considerably because of the opening of an intersubband scattering channel.
We calculate the electron density in the QW and in the parallel continuum channel as a function of the gate voltage, which shows a jump in capacitance at a 2DEG density corresponding to the occupation of the second subband ($E_2 < E_F$), in agreement with experiments~\cite{Manfra_100_InAs:2023} (cf.  Fig.~\ref{fig:density_voltage}).
We also predict the metal-insulator transition (MIT) in the parallel channel where bulk electrons near the Si doping layer breaks into puddles separated by random Coulomb disorder potential (cf. Table~\ref{table:samples}).
An experimental puzzle in InAs QWs is the mobility drop at an electron density lower than expected for the bound-state second subband occupation.
We explain this puzzle as a result of intersubband scattering between a bound state and a strongly level-broadened continuum resonance, mediated by screened Coulomb impurities [cf. Fig.~\ref{fig:mobility_density} (a)].
This level broadening effect is directly related to hybridization with continuum and to the single-particle scattering time $\tau_q$ through the quantum uncertainty principle $\Gamma \sim \hbar/\tau_q$, which characterizes the disorder and reflects QW quality. 

\textit{BSCR transition}.---The energy levels and the electron density profile in a conduction band modified by the electrostatic potential are obtained by solving the self-consistent Hartree equation at $T=0$~\cite{supplement_info}.
In our calculation, we use the same material parameters as the InAs sample reported in Ref.~\cite{Manfra_100_InAs:2023} [cf. Fig.~\ref{fig:quantum_well} (d), where $d_o=18$ nm, $d_1 = 10$ nm, $w=7$ nm, $d_2 = 4$ nm, and $d_d = 15$ nm].
(Because of the small InAs electron effective mass, exchange-correlation effects are negligible which we have explicitly checked by carrying out LDA calculations).
In addition to a constant electric field contributed from the metallic gate, the positively charged donors in the modulation doped Si $\delta$-layer of concentration $n_d =0.8\times 10^{12}$ cm$^{-2}$ attract electrons and pull down the conduction band edge, while the repulsive self energy of electrons pushes up the conduction band edge.
As a result, the QW splits into two conduction channels: the 2DEG channel of a concentration $n_{\mathrm{2DEG}}$ where electrons are localized inside the InAs QW, and the parallel continuum channel of a concentration $n_{\mathrm{para}}$ where electrons are distributed in the bulk centered near the Si doping layer. 
Meanwhile, the applied gate voltage $V_g$ pumps electrons in/out the QW by shifting the semiconductor Fermi level $E_{F,\mathrm{s}}$ relative to the Fermi level in the metallic gate $E_{F,\mathrm{metal}} =  E_{F,\mathrm{s}} - e V_g$.
In the rest of this letter, we use $E_F$ to represent $E_{F,\mathrm{s}}$.
Fig.~\ref{fig:quantum_well} (c) shows the calculated eigenenergies $E_l$ relative to $E_F$ as a function of $n_{\mathrm{2DEG}}$.
We see that the second bound state $E_2$, originally isolated from the continuum at high density, is pushed into the continuum and becomes a resonance by lowering $n_{\mathrm{2DEG}}$.
During this BSCR transition, the second subband occupation $E_2 <E_F$ is marked by a red dashed line.
Fig.~\ref{fig:quantum_well} (b) shows the conduction band edge $E_c(z)$ and the two-channel electron density profile (blue and red represent the 2DEG and parallel channel, respectively) corresponding to two different $n_{\mathrm{2DEG}}$ slices in Fig.~\ref{fig:quantum_well} (c).
We see that at small $n_{\mathrm{2DEG}} = 10\times 10^{11}$ cm$^{-2}$, only the first subband is occupied as a bound state, while the unoccupied second subband resides in the continuum as a resonance.
The resonance tunnels through the Schottky barrier $V_{\max}$ created by the Si donors with a finite probability, and hybridizes with the bulk continuum.
The tunneling action is given by $S = x_t \sqrt{2m (V_{\max} - E)}/\hbar$. 
For a tunneling barrier $(V_{\max} - E) \sim 30$ meV and tunneling length $x_t \sim 10$ nm, we have $S\sim 1$, so that the probability density inside the QW is $e^{S} \sim 3$ times larger than in the bulk.
The smaller $n_{\mathrm{2DEG}}$, the closer the resonance energy level to the top of the Schottky barrier with a smaller tunneling action, and the larger the hybridization with the continuum.
This hybridization opens a gap in the subband spectrum that decreases as the density increases.
On the other hand, at a high density $n_{\mathrm{2DEG}} = 25\times 10^{11}$ cm$^{-2}$, both the first and second subbands are occupied as bound states.
The second subband wavefunction overflows from the InAs QW to the InGaAs regime with a barrier height $V_1 = 180$ meV, confined by the higher InAlAs barrier with $V_2 = 300$ meV.

Below we analytically prove that InAs QW at low density is so shallow that only the lowest subband is a bound state while all excited subbands are in the continuum.
Since the wavefunction shape of the lowest subband does not change much by varying voltage at low density, we can use the finite potential well approximation to analytically compute the first subband energy $E_1$ by solving $\tilde{E}_1 + \tilde{E}_1 \tan^2(\sqrt{\tilde{E}_1}\pi/2) = \tilde{V}_1$~\cite{davies:1997,huang2023understanding}.
Here $\tilde{E} = E/E_1^{(\infty)}$ and $E_1^{(\infty)} = \hbar^2 \pi^2/2m w^2$ is the first subband energy assuming an infinite potential well.
For the InAs QW of interest, $E_1^{(\infty)} = 330$ meV, much larger than the InGaAs barrier $V_1 = 180$ meV, which justifies the shallow QW picture.
In the limit of $\tilde{V}_1 \ll 1$, we find $\tilde{E}_1 \approx (2/\pi^2) [-1+(1+\pi^2 \tilde{V}_1)^{1/2}]$.
Here $\tilde{V}_1 \approx 0.5$ and we obtain $\tilde{E}_1 \approx 0.3$.
Since $E_2 >E_1^{(\infty)}$ must be satisfied for the second subband, all excited resonances are in the continuum.

\begin{table}[b]
\vspace{-0.2 in}
\caption{Samples of different doping $n_d$ in Ref.~\cite{Manfra_100_InAs:2023} measured at $\mu_{\max}$. Physical quantities include the experimental (theoretical) parallel channel density $n_{\mathrm{para}}^{(\mathrm{exp})}$ ($n_{\mathrm{para}}$), the predicted MIT critical density $n_{c,\mathrm{para}}$ in units of $10^{11}$ cm$^{-2}$, and the experimental (theoretical) parallel mobility $\mu_{\mathrm{para}}^{(\mathrm{exp})}$ ($\mu_{\mathrm{para}}$) in units of $10^3$ cm$^{2}$/Vs. }
\begin{ruledtabular}
\begin{tabular}{c | c | c | c | c | c | c | c } 
Sample& $n_d$ & $n_{\mathrm{2DEG}}$& $n_{\mathrm{para}}^{(\mathrm{exp})}$ & $n_{\mathrm{para}}$ & $n_{c,\mathrm{para}}$ &  $\mu_{\mathrm{para}}^{(\mathrm{exp})}$ & $\mu_{\mathrm{para}}$ \\
\hline \\ [-3.2ex]
B & 8 & 7 & 0 & 1.2 & 2.2 & 0 & 1.8 \\
\hline \\ [-3.2ex]
C & 10 & 8 & 2.3 & 2.7 & 2.6 & 4.8 & 3.1 \\
\end{tabular}
\end{ruledtabular}
\label{table:samples}
\end{table}

\textit{Capacitance jump and the parallel MIT.}---The capacitance per unit area is defined as $C=edn_e/dV_g$.
In Fig.~\ref{fig:density_voltage} (a), we compute the electron density versus voltage by identifying $E_F = eV_g$ and fixing $E_{F,\mathrm{metal}}=0$ as the reference potential~\cite{supplement_info}, in good agreement with the experimental data (blue circles).
We see a capacitance jump at $n_{\mathrm{2DEG}} = 13\times 10^{11}$ cm$^{-2}$, corresponding to the second subband occupation.
This capacitance jump can be understood by a simple electrostatic model consisting of two capacitors in series $C^{-1}_{\mathrm{2DEG}} = C_g^{-1} + C_q^{-1} = 4\pi (d_{w} + d_q)/\kappa$.
Here $C_g = \kappa/4\pi d_{w}$ and $C_q = \kappa/4\pi d_q$ are the 2DEG geometric and quantum capacitance, respectively.
$\kappa = 15$ is the dielectric constant in InAs, $d_w = d_o + d_1 + w/2 = 31.5$ nm is the distance from the gate to the center of QW, $d_q = \kappa/2\pi e^2\nu$, and $\nu$ is the density of states (DOS).
Due to the second subband occupation, either as a bound state or as a continuum resonance, $\nu$ jumps by a factor of 2 and $d_q$ decreases from $35$ nm to $17$ nm, so that the effective capacitor thickness $(d_{w} + d_q)$ changes by a factor of $1.4$.
This explains that the 2DEG capacitance jumps up by a factor of $\sim 1.5$.
On the other hand, the parallel channel  becomes less sensitive to $V_g$ since the gate electric field is more effectively screened by 2DEG electrons with higher DOS.
Therefore, $C_{\mathrm{para}}$ jumps down while $C_{\mathrm{2DEG}}$ jumps up.

We predict a MIT in the parallel channel near the capacitance jump [cf. Fig.~\ref{fig:density_voltage} (a) and Table~\ref{table:samples}].
At small $n_{\mathrm{para}} \ll n_d$, electrons are localized and spatially separated as puddles in the parallel channel by random Coulomb disorder potential $\Gamma \sim \ln(\sqrt{n_d} a_B) e^2\sqrt{n_d}/\kappa$~\cite{supplement_info}, where $a_B=\kappa \hbar^2/m e^2 = 35$ nm is the effective Bohr radius.
When $E_F > \Gamma$ or $n_{\mathrm{para}} > n_{c,\mathrm{para}} \sim \ln(\sqrt{n_d} a_B) \sqrt{n_d}/a_B$, there is a MIT where electron puddles percolate through the disorder potential~\cite{Thouless:1971,Shklovskii:1972,Kirkpatrick:1973,Shklovskii:1975,shklovskii1984,DasSarma:2005579,Shklovskii:2007,Shaffique:2007,Manfra:2007,Tracy:2009,DasSarma:2005,Qiuzi:2013,Tracy:2014,Huang:2021a,Huang:2021,Huang:2022,Huang:2023}.
The parallel channel transitions from an insulator to a metal through the percolative 2D MIT when $n_{\mathrm{para}}$ increases by increasing $V_g$ or $n_d$.
In Table~\ref{table:samples}, we estimate the metallic parallel mobility through $\mu_{\mathrm{para}} = (e/\pi^2 \hbar n_d) (2k_{F}/q_{TF})^2$~\cite{supplement_info,Hwang_density_scaling:2013,huang2023understanding,huang2024electronic}, in good agreement with experiments~\cite{Manfra_100_InAs:2023}. 
Here $k_F=\sqrt{2\pi n_{\mathrm{para}}}$ and $q_{TF} = 2/a_B$ is the Thomas-Fermi screening wave vector.

Below we estimate the electron density at which second subband is occupied.
The QW depth $U=V_1$ is modified by the electric field $E = 4\pi e(n_{\mathrm{total}} - n_d)/\kappa$ across a distance $(d_1 + w/2)$ between the QW center and the oxide-semiconductor interface.
According to quantum mechanics, the necessary condition for the existence of a second bound state inside QW is $U > E_1^{(\infty)}$~\cite{davies:1997}.
Since $E_F$ is essentially pinned near the bottom of the continuum 
due to screening of the gate electric field by 2DEG, $E_F$ relative to the continuum subbands changes slowly when we change the electron density. 
Therefore, a simple estimation of the electron density when the second subband occupation is given by
$U=V_1 + E (d_1 +w/2)=E_1^{(\infty)}$.
Using $n_d = 8\times 10^{11}$ cm$^{-2}$, and $d_1 +w/2 = 13.5$ nm, we obtain $n_{\mathrm{total}} = 17 \times 10^{11}$ cm$^{-2}$, in agreement with the result shown in Fig.~\ref{fig:density_voltage}. 
We also check the condition $U=E_1^{(\infty)}$ estimates the second subband occupation density reasonably well for other samples with different doping densities~\cite{supplement_info,Manfra_100_InAs:2023}.

\textit{Mobility drop.}---Before we discuss the mobility drop in a shallow InAs QW~\cite{Manfra_100_InAs:2023} at high 2DEG density, 
we discuss the low-density mobility behavior where only the first subband is occupied as a bound state to extract disorder information.
Using Boltzmann transport theory~\cite{Ando_review:1982,huang2023understanding}, in Fig.~\ref{fig:mobility_density} (a), we compute the mobility versus density with realistic disorder scattering modeled as follows.
We model a $\delta$-layer of charged impurities of 2D concentration $n_{r}$ located at the semiconductor-oxide interface at $z=d_1+w/2 = 13.5$ nm away from the center of the QW.
Since $k_F z \gg 1$, they scatter 2DEG as remote impurities (RIs) with a characteristic mobility density dependence $\mu \propto n^{1.5}$~\cite{Hwang_density_scaling:2013}.
The other contribution to the RI scattering is from Si $\delta$-layer with a doping density $n_d = 0.8 \times 10^{12}$ cm$^{-2}$, located inside the bottom InAlAs barrier at $z = -(d_d + d_2 +w/2) = -22.5$ nm.
However, since the Si doping layer is further away from the center of the QW, the scattering from Si donors should contribute less than from the closer oxide RIs, as we confirm numerically in Fig.~\ref{fig:mobility_density}.

We model a uniform distribution of background charged impurities (BIs) with a 3D density $N_b$.
One can estimate $N_b$ from the mobility density data of buried InAs QWs reported in Ref.~\cite{Hatke:2017} [cf. Fig.~\ref{fig:mobility_density} (b), ``buried'' means that InGaAs barriers with $d_1,d_2>100$ nm are much thicker than those in shallow QWs], where a high peak mobility exceeding $10^6$ cm$^2$/Vs with $\mu \propto n^{\alpha}$ and $\alpha \sim$ 0.5--1~\cite{Shabani:2014,dempsey2024effectsstraincompensationelectron} suggests that the mobility of buried InAs QWs is limited by unintentional BIs with $N_b \sim 10^{15}-10^{16}$ cm$^{-3}$, in good agreement with the estimation reported in Ref.~\cite{Hatke:2017}.
Since the screening parameter $s=q_{TF}/2k_F = 0.36/\sqrt{(n_{\mathrm{2DEG}}/10^{11} \mathrm{cm}^{-2})}$ is much smaller than 1 in the experimental density range, the screening of BIs is rather weak, making them behave similar to RIs with a characteristic power law $\mu \propto n^{3/2}/\ln(s^{-1})$~\cite{Hwang_density_scaling:2013}. 
On the other hand, in shallow InAs QWs reported in Ref.~\cite{Manfra_100_InAs:2023}, the BI contribution to the mobility corresponding to $N_b = 1\times 10^{16}$ cm$^{-3}$ is much higher than the experimental mobility [cf. Fig.~\ref{fig:mobility_density} (a)], which suggests that BI is not the limiting scattering source.

In Fig.~\ref{fig:mobility_density}, we show that alloy disorder (AD) scattering is negligible in shallow InAs QWs which corresponds to a rather high mobility $\sim 4\times 10^6$ cm$^2$/Vs, while AD is one of the mobility limiting factors for buried InAs QWs. 
(In the calculation, we use the unscreened short-range AD potential $\Delta V = 0.65$ eV~\cite{Bastard:1983,Gold:1988,dong2024enhanced}).
Since the interfaces between InAs QW and InGaAs barriers should be of similar quality in Refs.~\cite{Manfra_100_InAs:2023} and \cite{Hatke:2017}, and since the interface roughness (IR) scattering predicts a mobility proportional to $w^6$~\cite{Gold:1988,Ando_review:1982}, the fact that the narrower $w=4$ nm buried QW~\cite{Hatke:2017} has a much higher mobility than the wider $w=7$ nm shallow QW~\cite{Manfra_100_InAs:2023} indicates that IR is not the limiting scattering mechanism in the wider QW (IR scattering is weaker for a wider QW).
The low-density quantum mobility $\mu_q$ is shown as a dot-dashed black curve in Fig.~\ref{fig:mobility_density} (a).
$\mu_q$ is much smaller than $\mu$ since small-angle forward scattering from distant charged impurity is included in $\mu_q$ but not in $\mu$~\cite{Stern:1985,Sammon:2018}. 

Next, we discuss the mobility drop at high density, which is a persistent prominent feature in shallow InAs QWs~\cite{Manfra_100_InAs:2023}.
Experimentally, mobility reaches the maximum $\mu_{\max}\sim 10^5$ cm$^2$/Vs at $n_{\mathrm{2DEG}} = 7\times 10^{11}$ cm$^{-2}$ [cf. Fig.~\ref{fig:mobility_density} (a)].
Such electron density is substantially smaller than expected for the second subband occupation $n_{\mathrm{2DEG}} \approx 12\times 10^{11}$ cm$^{-2}$ estimated from the capacitance jump (cf. Fig.~\ref{fig:density_voltage}), and from the Landau level (LL) crossing seen in magnetotransport~\cite{Manfra_100_InAs:2023}.
This is a serious conundrum.
We have two main reasons to believe that the best explanation for such a mobility drop is intersubband scattering between the bound-state first subband and the strongly level-broadened second subband as a continuum resonance, mediated by Coulomb impurities.
In other words, the level-broadened resonance effectively acts as a continumm with a continuous DOS tail that smears the mobility drop.
First, since the mobility continuously and smoothly drops for the whole higher density range, without showing other features, the only dominant scattering mechanism at high density should be intersubband scattering. 
Otherwise, we should have seen an additional sharp decrease in mobility when the second subband is occupied at $n_{\mathrm{2DEG}} \approx 12\times 10^{11}$ cm$^{-2}$, which is absent in the data.
Second, although short-range AD or IR usually affects high-density mobility stronger since the long-range charged impurities are screened~\cite{Hwang_density_scaling:2013,Hwang:2014_short_range}, we argue that AD and IR are negligible even if the second subband is occupied, and the dominant scattering source are charged impurities.
This is because the second subband wavefunction, residing in the InGaAs barrier, on average is much closer to the oxide and Si charged impurities compared to the first subband.
Consequently, Coulomb scattering is strongly enhanced and overwhelms other scattering sources.
For example, AD scattering rate is proportional to the squared probability leaking in the alloy region~\cite{Bastard:1983,Gold:1988,dong2024enhanced}. 
The leakage probability is $\sim 30$ \% for the first subband and $\sim 80$ \% for the second subband, leading to a decrease in mobility $\mu_{\mathrm{AD}} \sim 6\times 10^5$ cm$^2$/Vs.
However, such mobility is still much larger than the experimental mobility so AD is negligible.
IR scattering should not limits the second subband mobility either due to a larger effective well width $w_2=d_1+w+d_2\gg w$, contributing to a scattering rate $\propto w_2^{-6}$~\cite{Gold:1988,Ando_review:1982} even weaker than in first subband.

If there is no level broadening, there should be a sharp mobility decrease when the second subband is occupied~\cite{Mori_Ando:1979,Stormer:1982,Xie:1987,Fletcher:1988,Ensslin:1993,Muraki:2001} [cf. the solid black curve in Fig.~\ref{fig:mobility_density} (a)].
We approximate the multisubband 2DEG as a 3DEG distributed over a thickness of $w_2$ to estimate the high-density mobility, dominated by intersubband scattering mediated by Coulomb impurities, through the Brooks-Herring formula $\mu \approx 3\pi e a_B^2/4\hbar \ln(\pi k_{F} a_B)$~\cite{Brooks:1951,Brooks:1955,Chattopadhyay:1981,brooks_herring},
where $k_{F} \approx (3\pi^2 n_{\mathrm{2DEG}}/w_2)^{1/3}$.
As a result, the mobility at high density is $\sim 10^4$ cm$^2$/Vs, in reasonable agreement with experiments~\cite{Manfra_100_InAs:2023}.
However, the theory without level broadening predicts a sharper mobility drop at a larger density $n_{\mathrm{2DEG}} \approx 12\times 10^{11}$ cm$^{-2}$ compared to the smoother drop at $n_{\mathrm{2DEG}} = 7\times 10^{11}$ cm$^{-2}$ in experiments.
This discrepancy $\Delta n \approx 5 \times 10^{11}$ cm$^{-2}$ can be explained by a resonance strongly level-broadened via disorder and hybirdization with the continuum, so that the second subband resonance is smeared in an energy window of $\Gamma \sim \Delta n/g_0\sim 50$ meV.
The mobility $\mu = e\tau/m$ including the level broadening effect by smearing the scattering rate $\tau(\epsilon)^{-1}$ in an energy window of $\epsilon\pm \Gamma$ is shown as the red curve in Fig.~\ref{fig:mobility_density} (a)~\cite{coupled_boltzmann}.
Because of the long-range Coulomb potential fluctuation $\Gamma \sim (e^2 \sqrt{n_r}/\kappa) \ln(\sqrt{n_r} a_B)\approx 30$ meV, the second subband becomes an impurity band whose bottom is brought down by an amount of $\sim \Gamma$.
This is consistent with the broadening of LL crossings observed through magnetotransport~\cite{Manfra_100_InAs:2023}, where an energy-resolved LL crossing first shows up at $B\sim 6$ T corresponding to a cyclotron energy $\sim 30$ meV and to the second-subband quantum mobility $\mu_q=e\tau_q/m \sim 1/B \sim 10^3$ cm$^{2}$/Vs.
This is comparable to the low-density first-subband $\mu_q$ shown as a dot-dashed line in Fig.~\ref{fig:density_voltage} (a).
Similar results of $\Gamma=\hbar/\tau_q \sim 20$ meV have been measured from Dingle temperature of Shubnikov-de Haas oscillations in narrower InAs QWs~\cite{Yuan:2020}.
In addition, the hybridization between a resonance and the continuum also contributes to the level broadening.
The physics is that electrons can be scattered to subbands close to the resonance as long as their wavefunctions sufficiently localized inside QW that resemble excited ``bound states''.
At $n_{\mathrm{2DEG}} = 7\times 10^{11}$ cm$^{-2}$, the hybridization gap is $\sim 20$ meV [cf. Fig.~\ref{fig:quantum_well} (c)], comparable to $\Gamma$ induced by disorder.
The difference between the density corresponding to the capacitance jump and to the onset of mobility drop may be explained by a mobility edge $E_{\mu}$ near the level-broadened second subband bottom ($E_{\mu}\approx E_{2b}+\Gamma \approx E_2$).
Namely, the additional channel for scattering from the first to the second subband influences the mobility even though there are no mobile carriers in the second subband~\cite{Fletcher:1988}.
Similar level broadening effect has been discussed in GaAs~\cite{Stormer:1982,Fletcher:1988,Ensslin:1993,Muraki:2001}, but with a much smaller $\Gamma \lesssim 1$ meV corresponding to a much higher $\mu_q \gtrsim 10^4$ cm$^2$/Vs.

\textit{Conclusions.}---We have shown a transition from a bound state to a continuum resonance in a shallow quantum well, strongly tunnable by electrostatic gating. 
Using intersubband scattering involving a strongly level-broadened continuum resonance mediated by Coulomb impurities, we have explained the experimental puzzle in shallow InAs quantum wells that the mobility drops at an electron density smaller than expected for the bound-state second subband occupation.
Moreover, we have explained the experimentally observed capacitance jump and the metal-insulator transition in the parallel continuum channel accompanied the bound state-continuum resonance transition.
Our findings provide guidelines for enhancing the electrostatic control and sample quality of shallow InAs quantum wells, allowing further exploration and sample improvement of solid-state-based topological quantum computing platforms, which is currently the key stumbling block inhibiting the manifestation of non-Abelian Majorana zero modes.

We acknowledge helpful conversation with Alisa Danilenko.
This work is supported by the Laboratory for Physical Sciences.

\medskip


\begin{thebibliography}{64}
\bibitem [{bro()}]{brooks_herring}%
  \BibitemOpen
  \href@noop {} {}\bibinfo {note} {Although the charge compensation is unknown experimentally, we estimate the ratio of free carriers to Coulomb impurities as $n_{\mathrm{2DEG}}/(n_{\mathrm{2DEG}}+n_r) \approx 1/2$ at $n_{\mathrm{2DEG}}\approx 3\times 10^{12}$ cm$^{-2}$ in the Brooks-Herring formula.}\BibitemShut {Stop}%
\end{thebibliography}
\medskip

%


\setcounter{equation}{0}
\setcounter{figure}{0}
\setcounter{table}{0}
\makeatletter
\renewcommand{\theequation}{S\arabic{equation}}
\renewcommand{\thefigure}{S\arabic{figure}}
\renewcommand{\thetable}{S\arabic{table}}
\onecolumngrid
\section*{Supplementary materials}
\subsection{Self-consistent Hartree equation}
We discuss the algorithm we used to solve the self-consistent Hartree equation within the effective mass approximation at zero temperature $T=0$~\cite{Stern:1984}.
The 1D Poisson's equation for the electric potential $\phi(z)$ reads
\begin{align}
    \frac{d}{d z}\qty( \kappa(z) \frac{d \phi}{d z}) = -e (N_D - N_e[\phi]), \label{eq:poisson}
\end{align}
where $N_D =  n_d \delta(z-z_d) + n_o \delta(z-z_o) + n_g \delta(z-z_g)$ with $z_d$, $z_o$, and $z_g$ the positions of the Si $\delta$-doping layer, of the unintentional total charge in the oxide-semiconductor interface, and of the gate, respectively, while $n_d$, $n_o$, and $n_g$ are the corresponding 2D charge densities.
The charge neutrality condition ensures $n_d + n_o + n_g = n_e$, where $n_e = \int_{-\infty}^{+\infty} N_e(z) dz$ is the total electron density.
$\kappa(z)$ is the dielectric constant as a function of the position along $z$, which is material dependent.
The boundary conditions (the reference potential and the boundary electric field) for the Poisson equation Eq.~\eqref{eq:poisson} are
\begin{align}
    \phi(z=z_g) = 0\qc -\frac{d \phi}{d z} = \frac{4\pi e n_g}{\kappa_o},
\end{align}
where $\kappa_o$ is the dielectric constant of the gate dielectrics.
$N_e[\phi]$ is the 3D electron density given by 
\begin{align}\label{eq:Ne_phi}
    N_e[\phi] = \sum_{l=1}^{\infty} \Theta(E_F -  E_l) \abs{\psi_l(z)}^2 n_l, \\
    n_l = \frac{m_{\mathrm{DOS}}}{\pi \hbar^2} \int_{E_l}^{\infty} \Theta(E_F - \varepsilon) d\varepsilon,   \label{eq:n_l}
\end{align}
where $n_l$ is the electron density for the $l$-th subband, $m_{\mathrm{DOS}}/\pi \hbar^2$ is the density of states (DOS) for a subband with $m_{\mathrm{DOS}} = \sqrt{m_x m_y}$ the DOS effective mass of the 2DEG with a parabolic dispersion [c.f. Eq.~\eqref{eq:schrodinger_2DEG} below].
$E_l$ and $\psi_l(z)$ are the corresponding subband eigenenergy and wavefunction for the 1D Schr{\"o}dinger's equation along $z$-direction
\begin{align}\label{eq:schrodinger}
    -\frac{\hbar^2}{2m(z)} \frac{d^2 \psi_l(z)}{dz^2} +\qty[E_c(z) - e\phi(z)] \psi_l(z) = E_l \psi_l(z),
\end{align}
with a boundary condition $\psi_l(z \to \pm \infty) = 0$.
$E_c(z)$ is the conduction band bottom profile.
In deriving Eq.~\eqref{eq:schrodinger}, we assume $m_x$ and $m_y$ are independent on $z$, so that the full Schr{\"o}dinger's equation can be separated into two equations. 
The first is Eq.~\eqref{eq:schrodinger} which depends only on $z$, and the second is for the plane wave $\psi(x,y) \propto e^{ik_x x}e^{ik_y y}$ with a dispersion
\begin{align}\label{eq:schrodinger_2DEG}
    \varepsilon(k_x,k_y) = \frac{\hbar^2 k_x^2}{2m_x} + \frac{\hbar^2 k_y^2}{2m_y} = E_F- E_l,
\end{align}
and the total energy is $E_F$.
The electric potential $\phi$ enters Schr{\"o}dinger' equation as a Hartree energy, which in turn determines the eigenenergy and the electron density $N_e[\phi]$ that generate the electrostatic potential through Eqs.~\eqref{eq:Ne_phi} and \eqref{eq:n_l}. 
In other words, the electron density is a functional of the electric potential $\phi$, which makes the Poisson equation nonlinear.
As a result, Poisson's equation \eqref{eq:poisson} and Schr{\"o}dinger's equation \eqref{eq:schrodinger} should be solved together self-consistently.
\begin{figure*}
    \centering
    \includegraphics[width=\linewidth]{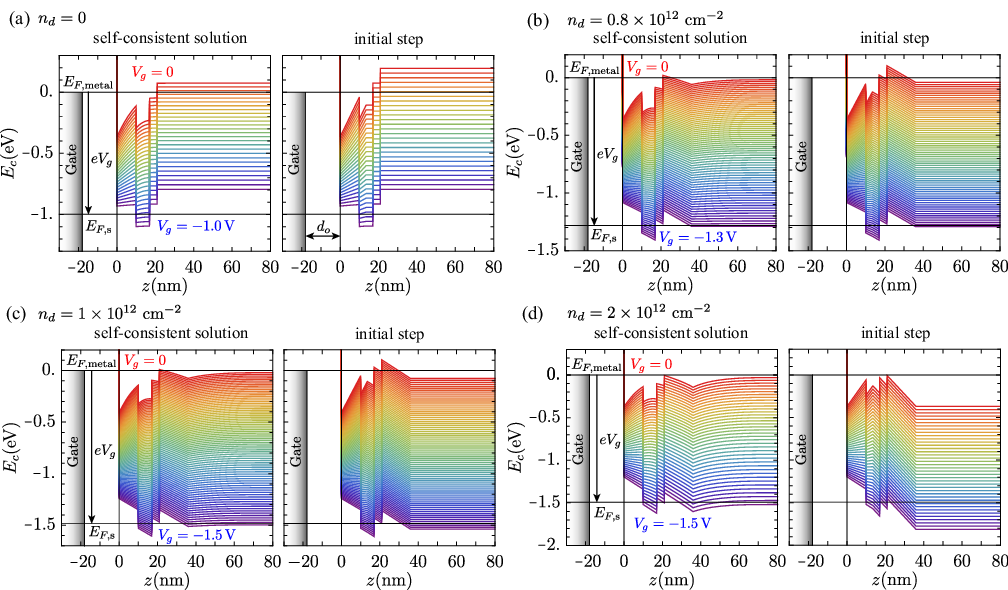}
    \caption{Self-consistent Hartree solution of the band edge profile $E_c$ for a range of voltage (colored curves). Samples A-D in Ref.~\cite{Manfra_100_InAs:2023} correspond to figure panels (a-d). At the initial step, all electrons are located at the QW center $N_e(z) = n_{\mathrm{total}} \delta (z-z_w)$.}
    \label{fig:Ec_Vg}
\end{figure*}

\begin{figure*}
    \centering
    \includegraphics[width=\linewidth]{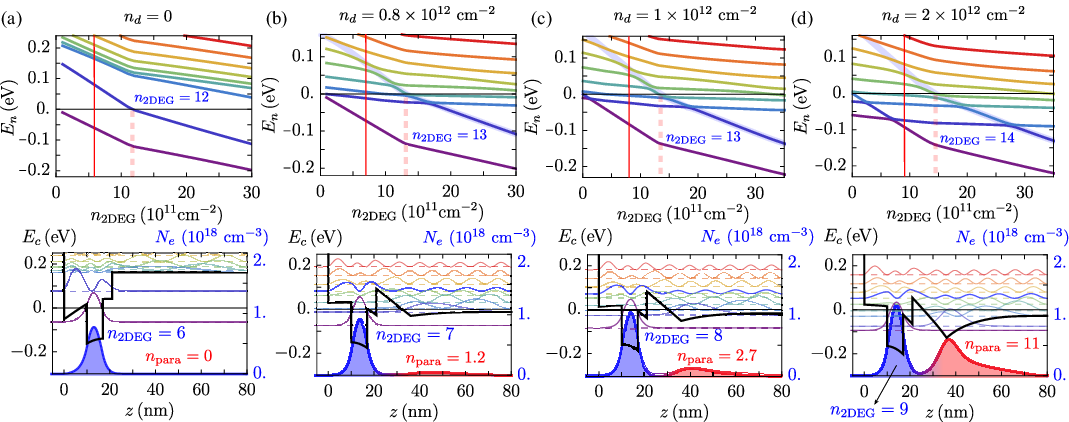}
    \caption{Samples A-D in Ref.~\cite{Manfra_100_InAs:2023} correspond to figure panels (a-d). The first row shows the subband energies $E_n$ versus $n_{\mathrm{2DEG}}$. The red dashed line marks the 2DEG density when the 2DEG second subband is occupied. The blue shadow line is a guide of the eye when the 2DEG second subband blurs into the continuum. The second row shows the conduction band edge $E_c$ (thick black) and the 3D electron density $N_e$ profile (blue crosses over to red) along $z$-direction.}
    \label{fig:Nd_En_Ec}
\end{figure*}

The self-consistent algorithm to solve Eqs.~\eqref{eq:poisson} and \eqref{eq:schrodinger} is as follows.
The initial state for the electron density we choose for the iterative algorithm is $N_{e,\mathrm{initial}} = n_e \delta(z-z_w)$ that generates the initial electric potential $\phi^{(0)}(z)$ according to the Poisson equation
\begin{align}
    \frac{d}{d z}\qty( \kappa(z) \frac{d \phi^{(0)}}{d z}) = -e (N_D - N_{e,\mathrm{initial}}). \label{eq:poisson_initial}
\end{align}
Given an initial $\phi^{(0)}(z)$, our goal is to find the exact solution to Poisson's equation, denoted $\phi(z)$. 
(We have checked that the convergent result of $\phi(z)$ after the iterative algorithm is robust in different initial conditions since the electrostatic solution is unique).
We achieve this goal by iteratively updating the electric potential, so that the sequence $\{\phi^{(k)}(z)\}$ converges to $\phi(z)$ in the sense that $\lim_{k \to \infty} [\phi^{(k+1)}(z) - \phi^{(k)}(z)] = 0$ for all positions $z$.
As the first step in the iteration, to obtain $\phi^{(1)}$ from $\phi^{(0)}$, we substitute $\phi^{(0)}$ for $\phi$ in Schr{\"o}dinger's equation Eq.~\eqref{eq:schrodinger} and obtain the corresponding electron density $N_e[\phi^{(0)}]$ through Eqs.~\eqref{eq:Ne_phi} and \eqref{eq:n_l}. 
We then substitute $N_e[\phi^{(0)}]$ into Eq.~\eqref{eq:poisson} as a source term, and we solve it for $\phi^{(1)}$.
In general, $\phi^{(k+1)}$ depends on $\phi^{(k)}$ through
\begin{align}\label{eq:iteration_poisson}
    \frac{d}{d z}\qty( \kappa(z) \frac{d \phi^{(k+1)}}{d z}) = -e (N_D - N_e[\phi^{(k)}]).
\end{align}
To ensure convergence of the iteration algorithm, we mix the electron density $N_e[\phi^{(k)}] \to \eta N_e[\phi^{(k)}] + (1-\eta) N_e[\phi^{(k-1)}]$ of the previous two adjacent steps before we substitute it into Poisson's equation to obtain the electric potential of the current step, so that $\abs{\phi^{(k+1)} - \phi^{(k)}} \ll \phi^{(k)}$ if we choose a small $\eta = 0.1$.
In other words, instead of Eq.~\eqref{eq:iteration_poisson}, we solve the following equation
\begin{align}
    \frac{d}{d z}\qty( \kappa(z) \frac{d \phi^{(k+1)}}{d z}) = -e (N_D - \eta N_e[\phi^{(k)}] - (1-\eta) N_e[\phi^{(k-1)}]).
\end{align}
We stop the iteration process at some finite $k$ once $[\phi^{(k)}(z) - \phi^{(k-1)}(z)]/\phi^{(k)}(z) < \delta$ for all positions $z$ with a small cutoff $\delta = 10^{-5}$, and the algorithm is numerically converged.

To model the conduction band bottom profile $E_c(z)$ for the device structures in Ref.~\cite{Manfra_100_InAs:2023} [c.f. Fig.~\ref{fig:quantum_well} (d) in the main text], we use the conduction band bottom difference $V_1 = 180$ meV between InAs and In$_{0.75}$Ga$_{0.25}$As, and $V_2 = 300$ meV between InAs and In$_{0.81}$Al$_{0.19}$As.
The difference in the work function between HfOx and In$_{0.75}$Ga$_{0.25}$As contributes to a high barrier $V_{o} = 2.7$ eV at the oxide-semiconductor interface.
We assume that there is a high barrier between the buffer layers and In$_{0.81}$Al$_{0.19}$As barrier, so that the wavefunction $\psi(z)$ vanishes at this boundary (we check that by lowering this barrier height, only the level spacing within the continuum states changes to a smaller value, which enhances the physics of the continuum described in the main text).
Since the effective masses in the In$_{0.75}$Ga$_{0.25}$As and In$_{0.81}$Al$_{0.19}$As barriers are not very different from the effective mass in InAs~\cite{Shabani:2018}, we use a uniform isotropic effective mass $m=0.023 m_0$ for the whole structure.
We choose a uniform dielectric constant $\kappa = 15$, since in HfOx $\kappa \sim 19$ is close to in semiconductor $\kappa \sim 15$.
The results of the conduction band bottom profile at different gate voltages for samples of different doping are shown in Fig.~\ref{fig:Ec_Vg}.
The corresponding subband energies $E_n$ relative to $E_F$ as functions of $n_{\mathrm{2DEG}}$ are shown in Fig.~\ref{fig:Nd_En_Ec} first row.

As explained in the main text, the gate electric field changes the electrostatic potential profile dramatically such that the portions of electron density populating the 2DEG and the parallel channels, $n_{\mathrm{2DEG}}$ and $n_{\mathrm{para}}$, are changed by the gate voltage.
Meanwhile, the gate voltage tune the total electron density by changing the position of the semiconductor Fermi level in the QW relative to the metal Fermi level in the gate (cf. Fig.~\ref{fig:Ec_Vg}).
As a result, $n_{\mathrm{2DEG}}$ and $n_{\mathrm{para}}$ are functions of the total electron density $n_e$.
A straight forward way to obtain the 2DEG density $n_{\mathrm{2DEG}}$ and the parasitic density $n_{\mathrm{para}}$ from the 3D electron density $N_e(z)$ is to first divide $N_e(z)$ into two parts by finding the minimum $z=z_m$ where $dN_e/dz=0$, so that 
\begin{align}
    n_{\mathrm{2DEG}} = \int_{-\infty}^{z_m} N_e(z) dz\qc n_{\mathrm{para}} = \int_{z_m}^{+\infty} N_e(z) dz.
\end{align}
Another way to compute $n_{\mathrm{2DEG}}$ and $n_{\mathrm{para}}$ is as follows.
We first find the dispersion of the subband energy as a function of the total electron density $E_l (n_e)$, and then identify the subbands that belong to the 2DEG channel $E_l^{(\mathrm{2DEG})}(n_e)$ by tracing the level crossings and doing the interpolation in $n_e$, such that $n_{\mathrm{2DEG}}$ is given by
\begin{align}
    n_{\mathrm{2DEG}} = \sum_{l=1}^{\infty} \Theta\qty(E_F -  E_l^{(\mathrm{2DEG})}) n_l^{(\mathrm{2DEG})}\qc n_l^{(\mathrm{2DEG})} = \frac{m}{\pi \hbar^2} \int_{E_l^{(\mathrm{\mathrm{2DEG}})}}^{\infty} \Theta(E_F - \varepsilon) d\varepsilon,
\end{align}
and $n_{\mathrm{para}} = n_e - n_{\mathrm{2DEG}}$.
We check that these two methods give essentially identical results of $n_{\mathrm{2DEG}}$ and $n_{\mathrm{para}}$, while the first method gives a slightly smoother dependence on $n_e$ because the second method requires a numerical interpolation, so we choose the first method to present our results. 

Figure~\ref{fig:Nd_En_Ec} second row shows the electron density profile (shaded color regime) and the conduction band edge (thick black curve) corresponding to a specific $n_{\mathrm{2DEG}}$ at $\mu_{\max}$ (cf. the red vertical line in Fig.~\ref{fig:Nd_En_Ec} first row).
We see that by increasing the doping density $n_d$, the electron density in the parallel channel increases dramatically while the increase in the electron density in the 2DEG channel is mediocre.
The thin solid (dashed) colored lines represent the wavefunction (eigenenergies).
For Sample A which is undoped, the second subband (thin blue) is a bound state overflowing to the InGaAs barrier regime, confined by the higher InAlAs barrier.
For Sample B-D which is modulation doped, the second subband in the 2DEG channel becomes an excited resonance hybridized with the continuum.

From Fig. 2 (a) reported in Ref.~\cite{Manfra_100_InAs:2023}, we notice that samples with different dopings all decay to the same mobility at high electron density, while the electron density at $\mu_{\max}$ increases as the doping density goes up [cf. Fig.~\ref{fig:Nd_En_Ec} second row].
The low-density mobility behavior are essentially the same for all samples of different doping.
This indicates that the scattering source responsible for the transport mobility are similar for different samples while the level broadening $\Gamma \sim \hbar/\tau_q$ are different for different samples due to the difference in the single-particle (quantum) scattering rate $\tau_q^{-1}$.
Since the quantum mobility $\mu_q = e\tau_q/m$ is affected by more distant Coulomb impurities which provide small-angle scattering while the transport mobility is more sensitive to the closer charged impurities that provide back scattering~\cite{Stern:1985,Sammon:2018}, this implies that the scattering amplitude of distant Coulomb impurities are different for different doping samples.
Indeed, for the undoped Sample A, there is no excess electrons in the bulk parallel channel to screen the distant charges, which decreases the quantum mobility and increases the level broadening.
For the doped Samples B-D, the excess electrons screen the distant charges more effectively and decreases the level broadening, which shifts the peak position closer to the second subband occupation.

\begin{figure*}
    \centering
    \includegraphics[width=\linewidth]{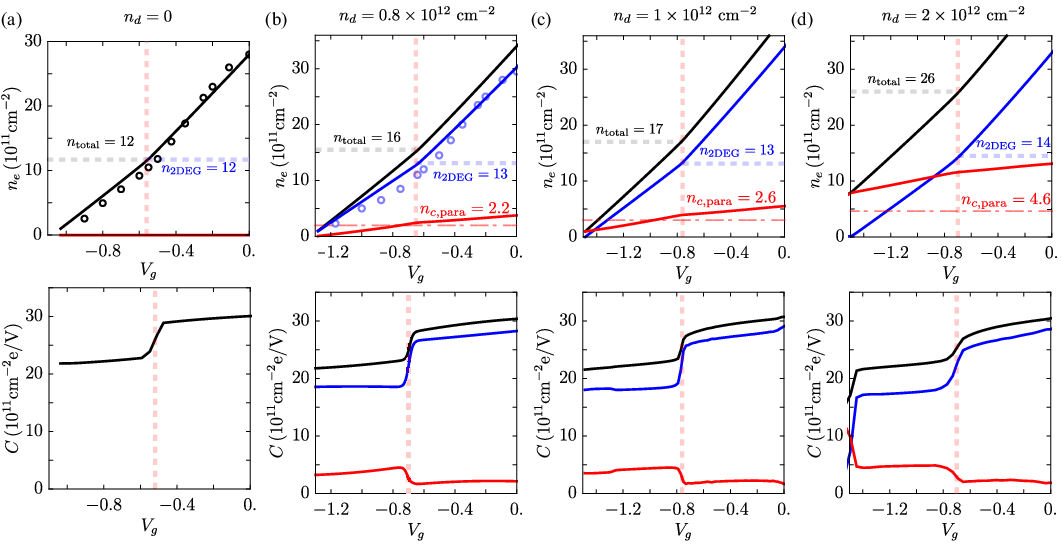}
    \caption{Samples A-D in Ref.~\cite{Manfra_100_InAs:2023} correspond to figure panels (a-d). The first row shows the densities $n_{\mathrm{total}}$ (black), $n_{\mathrm{2DEG}}$ (blue), and $n_{\mathrm{para}}$ (red) versus the gate voltage, where the zero voltage is calibrated by the experimental 2DEG density $n_{\mathrm{2DEG}}(V_g=0)=n_{\mathrm{2DEG}}^{(\mathrm{exp})}(V_g=0)$ and circles represent the data measured in Ref.~\cite{Manfra_100_InAs:2023}. The red horizontal dot-dashed line shows the MIT critical density for the parallel continuum channel. The second row shows the capacitance $C = e d n_e/d V_g$ for the total (black), 2DEG (blue), and the parallel (red) channel, respectively.}
    \label{fig:Nd_ne_Vg}
\end{figure*}

To compute the capacitance numerically, we assume the gate is a perfect metal with an infinite DOS, so that $E_{F,\mathrm{metal}}$ is not changed by applying gate voltage. 
We can use $E_{F,\mathrm{metal}}$ as a reference zero potential in the calculation of $E_{F,\mathrm{s}}$ by setting $E_{F,\mathrm{metal}} = 0$.
Using the relationship $E_{F,\mathrm{s}} = E_{F,\mathrm{metal}}+eV_g$, the differential capacitance per unit area is given by $C = edn_e/dV_g = -e^2dn_e/dE_{F,\mathrm{s}}$.
The results are shown in Fig.~\ref{fig:Nd_ne_Vg}.

\subsection{Metal-insulator transition}
Below we show that at the 2DEG density corresponding to $\mu_{\max}$ for samples of different Si doping densities in Ref.~\cite{Manfra_100_InAs:2023}, the parallel channels in Sample B have electron density smaller than the MIT critical density so they are localized and do not contribute to transport, while the parallel channels in Samples C and D have electron density larger than the MIT critical density so they are conducting, in agreement with Hall measurements in Ref.~\cite{Manfra_100_InAs:2023}.

The mean-squared fluctuation of potential energy generated by random Coulomb impurities in a 2D plane reads
\begin{align}\label{eq:gamma_squared}
    \Gamma^2 = n_d \int_{n_d^{-1/2}}^{r_s} d^2 r \qty(\frac{e^2}{\kappa r})^2 = 2\pi n_d \frac{e^4}{\kappa^2} \ln(n_d^{1/2} r_s).
\end{align}
Eq.~\eqref{eq:gamma_squared} is valid if the number of charged impurities $n_d r_s^2\gg 1$ inside the non-linear screening radius $r_s$ (which we justify below), and the charge number fluctuation is $\sqrt{n_d r_s^2}$. 
One can obtain the same result of mean-squared potential fluctuation in Eq.~\eqref{eq:gamma_squared} by calculating the on-site potential correlator $\ev{U(r) U(r)}$ in the Fourier space
\begin{align}
    \Gamma^2 = \ev{U(r) U(r)} = n_d \int \frac{d^2 q}{(2\pi)^2} \qty(\frac{2\pi e^2}{\kappa (q+ q_s)})^2\approx 2\pi n_d \frac{e^4}{\kappa^2}\ln(q_d/q_s),
\end{align}
where $q_d = \sqrt{n_d}$ is the ultraviolet cutoff and $q_s = 1/r_s$ is the screening wavevector.
Here $U(\vb{r}) = \sum_{\vb{R}_i} U_1 (\vb{r}-\vb{R}_i)$ is the total Coulomb potential summing all the impurities, $U_1(\vb{r} - \vb{R}) = (1/A) \sum_{\vb{q}} e^{i\vb{q}\cdot \vb{r}}(2\pi e^2/\kappa (q+q_s))$ is the Coulomb potential for a single Coulomb impurity located at $\vb{R}$, $A$ is the total area of the system, and $\ev{...} = (1/A) (\Pi_i \int d^2 R_i ...)$ averages over all the random realizations of impurities position.

When the kinetic energy of the parallel channel is smaller than the potential energy fluctuation $E_k = \hbar^2 \pi n_{\mathrm{para}}/m < \Gamma$, the electron density in the parasitic channel screens the charge fluctuation $\sqrt{n_d r_s^2}$ such that $n_{\mathrm{para}} = \sqrt{n_d r_s^2}/r_s^2 = \sqrt{n_d}/r_s$.
At the transition, the kinetic energy of electrons should be equal to the potential fluctuation so that electrons overcome the random potential barriers to percolate through the system and become delocalized~\cite{shklovskii1984}.
Solving $E_k = \Gamma$ together with $n_{\mathrm{para}} =\sqrt{n_d}/r_s$, we obtain
\begin{gather}\label{eq:n_c_para}
    n_{c,\mathrm{para}} \approx \frac{\sqrt{n_d}}{a_B} \sqrt{\frac{\pi}{8}} \ln(\sqrt{\pi n_d a_B^2/2}),\\
    r_{c,s} = n_d^{-1/2} \exp(\sqrt{\frac{\pi}{2n_d}} n_{c,\mathrm{para}} a_B) \approx n_d^{-1/2} (\pi n_d a_B^2/2)^{\pi/8}.\label{eq:nonlinear_rs}
\end{gather}
Using Eqs.~\eqref{eq:n_c_para} and \eqref{eq:nonlinear_rs}, the theoretically predicted $n_{c,\mathrm{para}}$ and $r_{c,s}$ for Samples B, C, and D with different Si doping densities are summarized in Table~\ref{table:parallel}.
We find $n_{c,\mathrm{para}} \sim 3\times 10^{11}$ cm$^{-2}$ and $r_{c,s} \sim a_B =35$ nm, which justifies the assumption that $n_{d} r_{s}^2 \sim 12 \gg 1$.
We mark these values of $n_{c,\mathrm{para}}$ as red dot-dashed horizontal lines in the middle row of Figs.~\ref{fig:Nd_ne_Vg} (b-d).
By comparing the predicted critical density with the carrier density in the parallel channel, we find the parallel channel in Sample B is insulating where $n_{\mathrm{para}} < n_{c,\mathrm{para}}$, while 
for Sample C and D the parallel channel is already metallic $n_{\mathrm{para}} >n_{c,\mathrm{para}}$.
We also estimate the mobility in the parallel channel through~\cite{Hwang_density_scaling:2013,huang2023understanding,huang2024electronic}
\begin{align}\label{eq:mobility_para}
    \mu_{\mathrm{para}} \approx \frac{e}{\pi^2 \hbar n_d} \qty(\frac{2k_{F}}{q_{TF}})^2,
\end{align}
where $k_{F} = \sqrt{2\pi n_{\mathrm{para}}}$, which is good for in-plane charged impurity scattering in the weak screening limit $q_{TF}/2k_F \ll 1$.
In deriving Eq.~\eqref{eq:mobility_para}, we assume there is a single subband in the parallel channel, which is good for Sample B and C, but not true for Sample D where there are two subbands occupided in the parallel channel as shown in Fig.~\ref{fig:Nd_ne_Vg} (d).
Similarly to the mobility drop in the 2DEG channel when the second subband is occupied, the multi-subband occupation in the parallel channel explains that the experimental parallel channel mobility for Sample D is smaller than the predicted value assuming a single subband (cf. Table~\ref{table:parallel}).
\begin{table}
\caption{ Samples of different doping $n_d$ in Ref.~\cite{Manfra_100_InAs:2023} measured at $\mu_{\max}$. Physical quantities include the experimental (theoretical) parallel channel density $n_{\mathrm{para}}^{(\mathrm{exp})}$ ($n_{\mathrm{para}}$), the predicted MIT critical density $n_{c,\mathrm{para}}$ [cf. Eq.~\eqref{eq:n_c_para}] in units of $10^{11}$ cm$^{-2}$, the non-linear screening length $r_{c,s}$ [cf. Eq.~\eqref{eq:nonlinear_rs}] in units of nm, the experimental (theoretical) parallel mobility $\mu_{\mathrm{para}}^{(\mathrm{exp})}$ ($\mu_{\mathrm{para}}$, cf. Eq.~\eqref{eq:mobility_para}) in units of $10^3$ cm$^{2}$/Vs, the ratio of the capacitance between the 2DEG and the parallel channel $C_{\mathrm{2DEG}}/C_{\mathrm{para}}$, and the residual density in the parallel channel when the 2DEG is completely depleted (cf. Eq.~\ref{eq:n_para_small_n_2DEG} and Fig.~\ref{fig:Nd_ne_Vg}).}
\begin{ruledtabular}
\begin{tabular}{c | c | c | c | c | c | c | c | c | c | c } 
Sample & $n_d$ & $n_{\mathrm{2DEG}}$& $n_{\mathrm{para}}^{(\mathrm{exp})}$ & $n_{\mathrm{para}}$ & $n_{c,\mathrm{para}}$ & $r_{c,s}$ & $\mu_{\mathrm{para}}^{(\mathrm{exp})}$ & $\mu_{\mathrm{para}}$ & $C_{\mathrm{2DEG}}/C_{\mathrm{para}}$ & $n_{\mathrm{para}}$ at $n_{\mathrm{2DEG}}=0$ \\
\hline \\ [-3.2ex]
A & 0 & 6 & 0 & 0 & --- & --- & 0 & 0 & --- & 0 \\
\hline \\ [-3.2ex]
B & 8 & 7 & 0 & 1.2 & 2.2 & 33 & 0 & 1.8 & 3.9 & 0 \\
\hline \\ [-3.2ex]
C & 10 & 8 & 2.3 & 2.7 & 2.6 & 32 & 4.8 & 3.1 & 3.8 & 1 \\
\hline \\ [-3.2ex]
D & 20 & 9 & 7.5 & 11 & 4.6 & 30 & 2.1 & 6.5 & 3.8 & 8
\end{tabular}
\end{ruledtabular}
\label{table:parallel}
\end{table}
\begin{table}
\caption{Critical densities $n_c$, $n_{cq}$, and the percolation threshold $n_{cp}$ in units of $10^{11}$ cm$^{-2}$. $n_c$ and $n_{cq}$ are estimated using the AIR criterion $k_F l = 1$ and $k_F l_q =1$. 
For 2D RIs $n_{cp} = 0.1 \sqrt{n_r}/d$, suitable for the shallow QW case, while $n_{cp} = (2/\sqrt{\pi})\sqrt{N_b/ a_B}$ for 3D BIs, suitable for the buried QW case. 
The results are calculated using the same impurity parameters from Fig.~\ref{fig:mobility_density} in the main text. }
\begin{ruledtabular}
\begin{tabular}{c | c | c | c} 
 & $n_c$ & $n_{cq}$ & $n_{cp}$ \\
\hline \\ [-3.2ex]
shallow QW~\cite{Manfra_100_InAs:2023} & 0.8 & 1.6 & 1.3\\
\hline \\ [-3.2ex]
buried QW~\cite{Hatke:2017} & 0.09 & 0.1 & 1.0
\end{tabular}
\end{ruledtabular}
\label{table:nc_air}
\end{table}

When $n_{\mathrm{2DEG}}$ is small, we can analytically calculate $n_{\mathrm{para}}$ using a simple electrostatic model.
The energy difference between $E_F$ and the bottom of the parallel continuum channel reads
\begin{align}\label{eq:delta_E_para}
    \Delta E \approx E_F(n_{\mathrm{2DEG}}) - \qty(V_0+V_1 - \frac{4\pi e^2 (n_d - n_{\mathrm{para}}) d_{w2}}{\kappa}),
\end{align}
where the parenthesis is the energy of the bottom of the continuum channel relative to the QW bottom, and $d_{w2} = d_d + d_2 + w/2$ is the distance between the doping layer and the QW center.
In Eq.~\eqref{eq:delta_E_para}, we drop the repulsive self-energy of electrons assuming their density is small.
The electron density in the parallel channel is given by 
\begin{align}\label{eq:n_para_small}
    n_{\mathrm{para}} \approx  g_{\mathrm{para}} \Delta E,
\end{align}
where $g_{\mathrm{para}}$ is the DOS of the parallel channel, and $g_{\mathrm{para}} = g_0=m/\pi \hbar^2$ if $n_{\mathrm{para}}$ is sufficiently small, while $g_{\mathrm{para}} = m k_{F,\mathrm{para}} L /\pi^2 \hbar^2$ if $n_{\mathrm{para}}$ is large enough so that many subbands are occupied.
Assuming that $n_{\mathrm{para}}$ is small enough such that only one parallel channel is occupied $g_{\mathrm{para}}=g_0$, and the 2DEG density is small enough such that $E_F = E_1 + \pi \hbar^2 n_{\mathrm{2DEG}}/m$, by solving Eq.~\eqref{eq:n_para_small} we obtain
\begin{align}\label{eq:n_para_small_n_2DEG}
    n_{\mathrm{para}} \approx \frac{n_{\mathrm{2DEG}} + 4 n_d d_{w2}/a_B - g_0 (V_0 + V_1 - E_1)}{1 + 4d_{w2}/a_B} \Theta(\Delta E),
\end{align}
where the Heaviside theta function $\Theta(\Delta E)$ makes sure $n_{\mathrm{para}} > 0$.
From Eq.~\eqref{eq:n_para_small_n_2DEG}, we can also predict the ratio of the capacitance in the 2DEG and the parallel channel
\begin{align}\label{eq:capacitance}
    \frac{C_{\mathrm{2DEG}}}{C_{\mathrm{para}}} = \frac{d n_{\mathrm{2DEG}}}{d n_{\mathrm{para}}} \approx 1 + 4d_{w2}/a_B.
\end{align}
Substituting the numbers from Ref.~\cite{Manfra_100_InAs:2023} into Eq.~\eqref{eq:capacitance}, we obtain $C_{\mathrm{QW}}/C_{\mathrm{para}} \approx 3.6$, in good agreement with the results shown in Fig.~\ref{fig:Nd_ne_Vg} and Table~\ref{table:parallel}.
Our theory correctly capture the MIT in the parallel channel and explain the corresponding parallel mobility measured in experiment. 
We also predict the MIT critical density in the 2DEG channel for Sample B in the shallow~\cite{Manfra_100_InAs:2023} and buried~\cite{Hatke:2017} InAs QWs using the Anderson-Ioffe-Regel criterion~\cite{anderson1958,ioffe1960} and the percolation threshold~\cite{Efros:19881019,Efros:1993,Huang:2021a,Huang:2022,Huang:2023}.
The result is shown in Table~\ref{table:nc_air}.


\end{document}